\documentclass[journal]{IEEEtran}

\usepackage{booktabs}
\usepackage{tabularx}
\usepackage{extarrows}
\usepackage{amssymb}
\usepackage{amsbsy}
\usepackage{amsmath}
\usepackage{bm}
\usepackage{verbatim}
\usepackage{cite}
\usepackage{mathrsfs}
\usepackage{amsfonts}
\usepackage{graphicx}
\usepackage[tight,footnotesize]{subfigure}
\usepackage[10pt]{moresize}
\usepackage{array}
\usepackage{color}
\usepackage{epsfig}
\usepackage{stfloats}
\usepackage{balance}
\usepackage{hyperref}  
\usepackage{amssymb} 
\usepackage[noend]{algpseudocode}

\usepackage{algorithmicx,algorithm}

\usepackage{cite}
\usepackage{setspace}
\usepackage{cases}
\usepackage{graphicx}
\usepackage{epstopdf}
\usepackage{multirow}
\usepackage{extarrows}
\newcommand{\subparagraph}{}
\usepackage{titlesec}
\usepackage{amsmath} %
\allowdisplaybreaks[4]

\ifCLASSINFOpdf

\else

\fi

\begin{document}
	\title{Using STAR-IRS to Secure Indoor Communications Through Symbol-Level Random Phase Modulation}

\author{Yanan~Du,~\IEEEmembership{Member,~IEEE,}
        Zeyang~Sun,~\IEEEmembership{Graduate Student Member,~IEEE,} 
        Yilan~Zhang, \\
        Sai~Xu,~\IEEEmembership{Member,~IEEE,}   
        Beiyuan Liu,~\IEEEmembership{Member,~IEEE}

 \thanks{This work was supported in part by the European Research Executive Agency's Horizon Europe MSCA 2022 Postdoctoral Fellowship CIRED under Grant 101109336 and in part by the Basic Research Programs of Taicang, 2023 under Grant TC2023JC24. 
 }

 \thanks{Yanan Du and Sai~Xu are with the School of Electrical and Electronic Engineering, University of Sheffield, Sheffield, S1 4ET, UK. (e-mail: yanan.du@sheffield.ac.uk, sai.xu@ieee.org).\par
 Zeyang Sun is with the School of Electronics and Information Engineering, Harbin Institute of Technology, Harbin, Heilongjiang, 150001, China (e-mail:zeyangsun97@gmail.com). \par
 Yilan~Zhang and Beiyuan~Liu are with the School of Cybersecurity, Northwestern Polytechnical University, Xi’an, Shaanxi, 710072, China (e-mail: 2023264604@mail.nwpu.edu.cn, lby@nwpu.edu.cn). }

\thanks{Manuscript received XX XX, XXXX; revised XX XX, XXXX. }} 
\maketitle
\begin{abstract}
This paper proposes a secure indoor communication scheme based on simultaneous transmitting and reflecting intelligent reflecting surface (STAR-IRS). Specifically, a transmitter (Alice) sends confidential information to its intended user (Bob) indoors, while several eavesdroppers (Eves) lurk outside. To safeguard the transmission from eavesdropping, the STAR-IRS is deployed on walls or windows. Upon impinging on the STAR-IRS, the incoming electromagnetic wave is dynamically partitioned into two components, enabling both transmission through and reflection from the surface. The reflected signal is controlled to enhance reception at Bob, while the transmitted signal is modulated with symbol-level random phase shifts to degrade the signal quality at Eves. Based on such a setting, the secrecy rate maximization problem is formulated. To solve it, a graph neural network (GNN)-based scheme is developed. Furthermore, a field-programmable gate array (FPGA)-based GNN accelerator is designed to reduce computational latency. Simulation results demonstrate that the proposed strategy outperforms both the conventional scheme and the reflection-only scheme in terms of secrecy performance. Moreover, the GNN-based approach achieves superior results compared to benchmark techniques such as maximum ratio transmission (MRT), zero forcing (ZF), and minimum mean square error (MMSE) in solving the optimization problem. Finally, experimental evaluations confirm that the FPGA-based accelerator enables low inference latency.
\end{abstract}
\IEEEpeerreviewmaketitle
\section{Introduction}
\IEEEPARstart{T}HE FORTHCOMING sixth-generation network is anticipated to deliver an unprecedented 1,000-fold increase in data traffic \cite{W. Saad}, driven primarily by the rapid proliferation of data-intensive applications such as large-scale machine-type communications, virtual reality, and Internet-of-Things devices. Given that most wireless data transmissions occur indoors \cite{Cisco, K. Doppler}, securing indoor wireless communications against eavesdropping threats has emerged as a critical and timely research challenge. Owing to the intrinsic openness of wireless propagation, transmissions within indoor environments are particularly vulnerable to interception by eavesdroppers located externally \cite{Z. Zhang}, posing substantial risks to information confidentiality. Consequently, there is an urgent need to develop security mechanisms tailored for indoor confidential transmissions to counter external eavesdropping.\par
Physical layer security (PLS) offers a compelling approach to safeguard wireless communications by exploiting the physical characteristics of channels, without relying solely on conventional cryptographic methods. Typical PLS strategies, including beamforming and artificial noise (AN), have been widely applied \cite{Z. Lin,C. Gong}. Secure beamforming concentrates signal energy towards intended users while suppressing or nullifying it towards eavesdroppers \cite{Z. Lin}. Meanwhile, AN introduces controlled interference to degrade eavesdroppers’ signal-to-interference-plus-noise ratios (SINRs), with minimal impact on intended users \cite{C. Gong}. However, little research has been conducted on external eavesdropping through physical barriers such as walls or windows.\par
A simultaneously transmitting and reflecting intelligent reflecting surface (STAR-IRS) operates as a sophisticated planar metasurface densely integrated with a multitude of tunable passive elements. These elements are designed to simultaneously divide and process incident electromagnetic energy, allowing one portion to be reflected back toward the source side while the other portion is transmitted through to the opposite side. Each element applies an independent adjustable coefficient to the reflected and transmitted components \cite{J. Xu}. Through joint configuration of all elements, the STAR-IRS can achieve 360° coverage, simultaneously forming beams in the reflection and transmission half-spaces. In essence, the STAR-IRS extends the coverage of IRS technology \cite{Xu2022Intelligent, Zhang2024Intelligent, Zhu2025Transmissive, Du2025Intelligent} to full-space, providing additional degrees of freedom (DoF) in shaping the wireless propagation environment. With these capabilities, a STAR-IRS deployed on walls or windows of an indoor venue can have an effect on both indoor and outdoor users concurrently, making it a promising solution for indoor scenarios where eavesdroppers might lurk just outside the facility. Importantly, STAR-IRS maintains the key advantages of IRS, including minimal power consumption and reduced hardware complexity, making it highly suitable for practical deployment in indoor scenarios. This unique capability of STAR-IRS motivates its extensive investigation in various wireless communication contexts, with a particular emphasis on enhancing indoor PLS.
\subsection{Prior Works}
To effectively enhance wireless security, it is essential to develop advanced optimization approaches that leverage the unique capabilities of STAR-IRS. This paper will investigate STAR-IRS aided indoor wireless communications for security and employ deep learning to optimize system performance. A review is provided on STAR-IRS-assisted PLS strategies as well as deep learning (DL)-based approaches for STAR-IRS-related optimization, respectively.
\subsubsection{STAR-IRS Aided PLS}
Niu \textit{et al.} \cite{H. Niu 2} explored the weighted sum secrecy rate maximization in a STAR-IRS assisted multiple-input single-output (MISO) network, considering three distinct transmission protocols: energy splitting, mode selection, and time splitting.  Wan \textit{et al.} \cite{Wan2024b} investigated a STAR-IRS aided multiple-input multiple-output (MIMO) network for physical layer key generation, developing a penalty-based algorithm to maximize the derived closed-form sum secret key rate. Similarly, Shen \textit{et al.} \cite{Shen_Power} proposed a penalty-based alternating optimization method to jointly design transmit covariance matrices and STAR-IRS coefficients, aiming to minimize power consumption in secure MIMO communications for perfect and imperfect channel state information (CSI). Han \textit{et al.} \cite{Y. Han} studied secure non-orthogonal multiple access communications with STAR-IRS, incorporating AN in transmitted signals to mitigate eavesdropping threats, optimizing secrecy rate under individual secrecy constraints and total transmit power limitations. Sun \textit{et al.} \cite{Sun2023} leveraged phase-coupled intelligent omnidirectional surfaces to enhance  integrated sensing and communications system security by minimizing information leakage to potential malicious entities while satisfying minimum signal-to-interference-plus-noise ratio (SINR) requirements. Additionally, Chi \textit{et al.} \cite{WPC} analyzed three secure transmission schemes within STAR-IRS assisted wireless powered communication networks under various blockage scenarios, providing closed-form outage and intercept probability expressions. However, research on STAR-IRS-enhanced indoor communications \cite{F. Yu,H. Chi} remains relatively limited.
\subsubsection{DL-based IRS/STAR-IRS Optimization}
DL-based optimization and resource allocation methods have been explored for IRS and STAR-IRS aided wireless communication systems \cite{Guo2023,Zhong2023,Zhang_Security}. Particularly, deep reinforcement learning (DRL) has demonstrated substantial potential for autonomously configuring a large number of passive elements.  Guo \textit{et al.} \cite{Guo2023} employed a deep deterministic policy gradient (DDPG) algorithm to jointly optimize the precoding vectors at the base station (BS) and the STAR-IRS coefficients, resulting in significant enhancements in energy efficiency. Zhong \textit{et al.} \cite{Zhong2023} proposed a hybrid DRL method combining DDPG and deep $Q$-learning, effectively minimizing long-term power consumption through joint active-passive beamforming optimization. Zhang \textit{et al.}  \cite{Zhang_Security} studied STAR-IRS-assisted secure MISO transmission and proposed DRL-based algorithms to jointly optimize beamforming and STAR-IRS coefficients, showing improved secrecy rate. However, these DRL approaches typically suffer from several drawbacks, such as the requirement for extensive training episodes, high computational complexity during training, and limited generalization capabilities. As a promising alternative, graph neural networks (GNNs) have emerged for wireless resource allocation tasks due to their intrinsic capability to model network topology and scale efficiently \cite{Chen2024distribute,GNN3}. Specifically, Chen \textit{et al.} \cite{Chen2024distribute} developed an IRS-enhanced cell-free MIMO network and used a distributed GNN algorithm to jointly optimize BS beamforming and IRS reflection.  Xu \textit{et al.}~\cite{GNN3} introduced a cluster-free multi-cell NOMA scheme using an auto-learning GNN approach to efficiently handle interference and reduce computation and communication overhead. However, no existing research has explored GNN-based joint beamforming optimization specifically in STAR-IRS assisted secure indoor communication settings.
\subsection{Motivations and Contributions}
Up to now, secure indoor communications leveraging STAR-IRS have not yet been extensively studied. Furthermore, the approach of symbol-level random phase modulation for generating AN at the STAR-IRS has not been considered. Motivated by this observation, this paper proposes to employ STAR-IRS to secure indoor communications through symbol-level random phase modulation. The main contributions can be summarized as follows: 
\begin{itemize}
	\item  A STAR-IRS-based secrecy strategy with symbol-level random phase modulation is first proposed for indoor communications. Specifically, the incident signal at the STAR-IRS is intelligently split into two distinct components: a reflected information-bearing signal directed towards the intended indoor user, and a transmitted AN signal for degrading potential outdoor eavesdroppers. This design significantly improves the reception quality for the intended user while concurrently impairing the signal reception capability of eavesdroppers.
	\item The secrecy rate maximization (SRM) problem is mathematically modeled under the total power budget constraint as well as the reflection and transmission coefficients constraints. To address the non-convex problem, a GNN-based deep learning algorithm is designed to simultaneously optimize the transmitter’s beamforming vector as well as the STAR-IRS’s transmission and reflection matrices. 
    \item  The trained GNN model is implemented on a field-programmable gate array (FPGA) board. By exploiting hardware parallelism and structural optimizations, the FPGA-based accelerator aims to minimize inference latency and power consumption to meet the stringent real-time demands of indoor communication systems. 
    \item Simulations and experiments validate the strong secrecy performance of the proposed STAR-IRS-based secure indoor communication strategy, demonstrate the superiority of the GNN-based optimization scheme over traditional methods, and confirm the low-latency capability of the FPGA-based GNN accelerator.
\end{itemize}
\par The remainder of this paper is organized as follows. Section II describes the system model and formulates the SRM problem. Section III details the proposed GNN-based solution for joint beamforming optimization. Section IV develops the FPGA-based accelerator. Section V presents simulation and experimental results. Section V concludes this paper.
\section{System Model and Problem Formulation}
\subsection{System Model}
\begin{figure*}
\centering
\includegraphics[width=.99\linewidth]{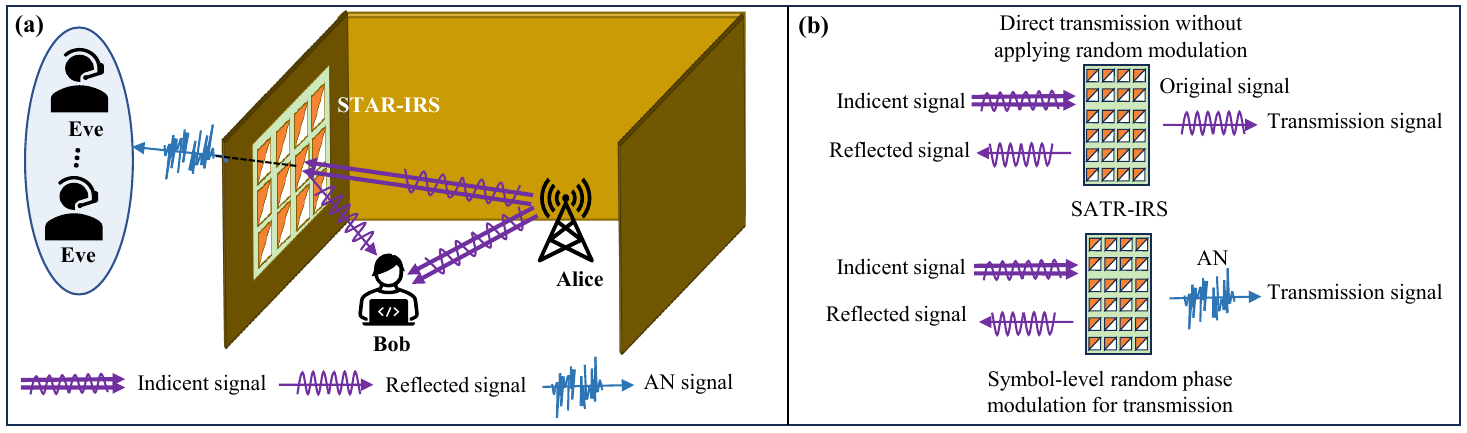}
\caption{(a) An illustration of the STAR-IRS-based secure indoor communication system. (b) A comparison between direct transmission without symbol-level random phase modulation and with it.}\label{Fig1}
\end{figure*}

As illustrated in Fig. \ref{Fig1}(a), a STAR-IRS-based secure indoor communication system is considered. In this system, an $N$-antenna base station (Alice) transmits confidential information to an intended single-antenna indoor user (Bob). Meanwhile, $K$ single-antenna eavesdroppers (Eves), indexed by $k \in \mathcal{K} = \{1,2,\dots,K\}$, are passively located outdoors, attempting to intercept the transmitted signal. To safeguard the communication against eavesdropping, a STAR-IRS, comprising $L$ passive elements, is deployed on the interior walls or windows. Unlike the conventional IRS that only reflects incident signals, a STAR-IRS enables each element to perform both reflection and transmission. This dual functionality allows the incident signal to be split into two components: one reflected toward Bob and the other transmitted, potentially reaching Eves.\par
This unique characteristic enables enhanced physical-layer security by exploiting the spatial separation between Bob and Eves. A comparison of traditional pure reflective STAR-IRS and the proposed symbol-level random phase modulation based STAR-IRS is depicted in Fig. \ref{Fig1}(b). Specifically, the reflected signal carries the confidential information to reinforce the signal reception at Bob. On the other hand, in contrast to prior works (e.g., \cite{H. Niu 2, Y. Han}) where direct signal transmission is employed, we consider a novel approach in which symbol-level random phase modulation is applied to the transmitted signal. This modulation transforms the signal into AN, which effectively degrades the reception quality at Eves, thereby strengthening secrecy. \par
Assume that wireless channels experience quasi-static flat fading, ensuring that the channel state information (CSI) remains invariant throughout each coherence interval. Define $\textbf{h}_\text{b} \in \mathbb{C}^{N \times 1}$ and $\textbf{h}_k \in \mathbb{C}^{N \times 1}$ as the direct channel vectors from Alice to Bob and the $k$-th Eve, respectively. Let $\textbf{f}_\text{b} \in \mathbb{C}^{L \times 1}$ and $\textbf{f}_k \in \mathbb{C}^{L \times 1}$ denote the channel vectors from the STAR-IRS to Bob and the $k$-th Eve, respectively. $\textbf{G} \in \mathbb{C}^{L \times N}$ denotes the channel matrix from Alice to the STAR-IRS. Denote $\textbf{w} \in \mathbb{C}^{N \times 1}$ as the beamforming vector corresponding to the confidential signal $s$ with $\mathbb{E}[|s|^2] = 1$. Based on the considered model, the signals received at Bob and the $k$-th Eve are respectively expressed as
\begin{align}
&y_\text{b}=\textbf{h}^H_\text{b} \textbf{w} s + \textbf{f}^H_\text{b} \boldsymbol{\Omega}_\text{r} \textbf{G}\textbf{w} s + n_\text{b}, \nonumber
\end{align}
and
\begin{align}
&y_k = \textbf{h}^H_k \textbf{w} s +\textbf{f}^H_k \boldsymbol{\Omega}_\text{t} \textbf{G}\textbf{w} s + n_k, ~ k \in \mathcal{K}, \nonumber
\end{align}
where $\boldsymbol{\Omega}_\text{r}$ and $\boldsymbol{\Omega}_\text{t}$ denote the diagonal reflection and transmission coefficient matrices of the STAR-IRS, respectively. The additive noise terms $n_\text{b} \sim \mathcal{CN}(0,\sigma_\text{b}^2)$ and $n_k \sim \mathcal{CN}(0,\sigma_k^2)$ denote the complex Gaussian noise at Bob and the $k$-th Eve, respectively .\par
Each STAR-IRS element supports simultaneous reflection and transmission, with associated complex coefficients. Specifically, for the $l$-th STAR-IRS element ($l \in \mathcal{L} = \{1,2,\dots,L\}$), the reflection and transmission coefficients are defined as $\sqrt{\beta^\text{r}_l} e^{j\theta^\text{r}_l}$ and $\sqrt{\beta^\text{t}_l} e^{j\theta^\text{t}_l}$, respectively, where $\beta^\text{r}_l \in [0,1]$ and $\beta^\text{t}_l \in [0,1]$ represent the reflection and transmission energy coefficients of the $l$-th element, respectively. The energy conservation principle imposes the constraint $\beta^\text{r}_l + \beta^\text{t}_l = 1$, while the phase shifts $\theta^\text{r}_l, \theta^\text{t}_l \in [0,2\pi)$ can be independently adjusted. Additionally, $\theta^r_l\in [0,2\pi)$ and $\theta^t_l \in [0,2\pi)$ denote the phase shifts of the $l$-th element, respectively. It is important to note that $\beta^\text{r}_l$ and $\beta^\text{t}_l$ are coupled due to the law of energy conservation, whereas $\theta^\text{r}_l$ and $\theta^\text{t}_l$ can be adjusted independently.\par
In the proposed STAR-IRS-aided secure communication strategy, the reflected signal is bounced back directly without processing, while the transmitted signal undergoes random phase changes at the symbol level, transforming into AN during transmission. Mathematically, the modulation process into AN is formulated as 
\begin{equation} \label{eq:2}
\textbf{f}^H_k \boldsymbol{\Omega}_\text{t} \textbf{G}\textbf{w} s 
  = \textbf{f}^H_k \tilde{\boldsymbol{\Omega}}_\text{t} \check{\boldsymbol{\Omega}}_\text{t} s  \textbf{G}\textbf{w} 
    = \textbf{f}^H_k \tilde{\boldsymbol{\Omega}}_\text{t} \textbf{Z}  \textbf{G}\textbf{w}  = \textbf{f}^H_k \tilde{\boldsymbol{\Omega}}_\text{t}  \textbf{G}\textbf{w} z, \nonumber
\end{equation}
where $\check{\boldsymbol{\Omega}}_\text{t}$ is defined as $\check{\boldsymbol{\Omega}}_\text{t} = \textup{diag}(e^{j\check{\theta}_1},e^{j\check{\theta}_2},...,e^{j\check{\theta}_L})$. $\check{\boldsymbol{\Omega}}_\text{t}$ is used to transform $s$ into AN $z$ by rapid and random change of the phases $\check{\theta}_1$, $\check{\theta}_2$, ..., $\check{\theta}_L$ at the symbol level. $\textbf{Z} = \check{\boldsymbol{\Omega}}_\text{t} s$ denotes the AN matrix. For simplicity of control, $\check{\theta}_1 = \check{\theta}_2 = \cdots = \check{\theta}_L$ is set. Hence, $\textbf{Z} = \boldsymbol{I} z$, where $z$ denotes AN. $\tilde{\boldsymbol{\Omega}}_\text{t}$ represents the diagonal transmission coefficient matrix for AN passive beamforming, where the $l$-th element is denoted as $\sqrt{ \beta^\text{t}_l}e^{j \tilde{\theta}^\text{t}_l}$.  \par
Based on this, the signal received at the $k$-th Eve can be rewritten as
\begin{align}
y_k = \textbf{h}^H_k \textbf{w} s + \textbf{f}^H_k \tilde{\boldsymbol{\Omega}}_\text{t}  \textbf{G}\textbf{w} z + n_k, ~ k \in \mathcal{K}. \nonumber
\end{align}
Accordingly, the SINRs at Bob and the $k$-th Eve are respectively given by
\begin{align}
&\gamma_\text{b} = \frac{ | (\textbf{h}_\text{b}^H+\textbf{f}_\text{b}^H \boldsymbol{\Omega}_\text{r} \textbf{G}) \textbf{w} |^2}{\sigma_\text{b}^2}, 
\end{align}
and
\begin{align}
&\gamma_{k} =\frac{ | \textbf{h}_k^H\textbf{w}|^2}{ |\textbf{f}_k^H\tilde{\boldsymbol{\Omega}}_\text{t}\textbf{G} \textbf{w} |^2 + \sigma_k^2 },~ k \in \mathcal{K}. \label{eq2}
\end{align}
Based on these expressions, the achievable secrecy rate is defined as
\begin{equation} \label{eq:4}
R_{s} = \left[ \log_2\left ( 1 + \gamma_\text{b} \right) - \max_{\forall k \in \mathcal{K}} ~\log_2\left ( 1 + \gamma_k \right) \right]^+, \nonumber
\end{equation}
where $[x]^+=\text{max}\{x,0\}$ ensures non-negativity of the secrecy rate.\par
\subsection{Problem Formulation}
This work aims to enhance the secrecy rate for an indoor communication system incorporating a STAR-IRS, through the joint design of the active beamforming vector $\textbf{w}$ at Alice,  the reflection coefficient matrix $ \boldsymbol{\Omega}_\text{r}$, and the transmission coefficient matrix $\tilde{\boldsymbol{\Omega}}_\text{t}$ at the STAR-IRS. Mathematically, the SRM problem is given by
\begin{equation}\label{P1}\nonumber
	\begin{aligned}
		(\text{P1})\quad \mathop{\max}_{\textbf{w}, \boldsymbol{\Omega}_\text{r},\tilde{\boldsymbol{\Omega}}_\text{t}} \quad & R_{s} \\
		s.t. \quad 
		&\text{C1}: \text{Tr}(\textbf{{ww}}^H) \leq P_{\text{max}}, \\
		&\text{C2}: \beta^\text{r}_l + \beta^\text{t}_l =1,\quad  \forall l\in \mathcal{L}, \\
  		&\text{C3}: 0\leq \beta^\text{r}_l \leq1,\quad  \forall l\in \mathcal{L},\\
    	&\text{C4}: 0\leq \beta^\text{t}_l \leq1,\quad  \forall l\in \mathcal{L},\\
		&\text{C5}: 0\leq  \theta^\text{r}_l \leq 2\pi,\quad  \forall l\in \mathcal{L},\\
  		&\text{C6}: 0\leq\tilde{\theta}^\text{t}_l\leq 2\pi,\quad  \forall l\in \mathcal{L},\\
	\end{aligned}
\end{equation}
where $P_{\text{max}}$ is the maximum transmit power at Alice, constraint C1 imposes a power budget on Alice's transmitted signals, C2 ensures the energy conservation principle at each STAR-IRS element, mandating the sum of reflection and transmission energy coefficients to be exactly one, C3 and C4 specify the permissible ranges for reflection and transmission energy coefficients, C5 and C6 regulate  the reflection and transmission phase shifts of STAR-IRS elements to vary continuously from $0$ to $2\pi$.
\section{GNN-Based Beamforming Scheme}
The problem (P1) is inherently challenging due to several factors: 1) The secrecy rate objective is defined by the difference of logarithmic SINR terms, making the optimization non-convex and difficult to manage analytically. This structure impedes the direct application of traditional convex optimization methods; 2) The optimization objectives involve enhancing Bob's signal strength while simultaneously suppressing signal quality at Eves. This adversarial objective further complicates the optimization. Additionally, the incorporation of symbol-level random phase modulation introduces dynamic AN, significantly increasing complexity in modeling and optimization. 3) The transmission and reflection coefficients of each STAR-IRS element are tightly coupled through the energy conservation constraint, resulting in a high degree of interdependence among optimization variables. Adjustments to one coefficient directly affect the other, thus complicating the optimization substantially. To address the problem (P1), we propose a GNN-based scheme to efficiently perform joint optimization of the optimization variables $\textbf{w}$, $\boldsymbol{\Omega}_{\text r}$, and $\tilde{\boldsymbol{\Omega}}_{\text t}$.
\subsection{Input}
To effectively utilize the GNN-based scheme for solving the optimization problem (P1), a graph representation must be created from the indoor communication network to serve as the input for the GNN. The graph is typically represented as \(\mathcal{G} = (\mathcal{V}, \mathcal{E})\), where \(\mathcal{V}\) and \(\mathcal{E}\) refer to the collection of nodes and edges connecting these nodes, respectively. In the optimization problem (P1), the direct and cascaded channels of Bob and Eves are interrelated, making them suitable to be represented as the node features in the graph.\par
Given that neural networks face challenges when handling complex-valued data directly, we separate the real and imaginary components of all relevant channels to construct meaningful and interpretable node features. Specifically, the Bob and Eve node feature vectors are formulated using their direct and cascaded channels, incorporating the real and imaginary components separately. For Bob and the $k$-th Eve, the feature vectors are defined as
\begin{align}
{{\textbf{x}}_\text{b}} &= {\left[ {{\text{Re}}\left( {{\textbf{h}}_\text{b}^T} \right), {\text{Im}}\left( {{\textbf{h}}_\text{b}^T} \right), {\text{Re}}\left( {{\textbf{d}}_\text{b}^T} \right), {\text{Im}}\left( {{\textbf{d}}_\text{b}^T} \right)} \right]^T}, \nonumber
\end{align} 
and
\begin{align}
{{\textbf{x}}_k} &= {\left[ {{\text{Re}}\left( {{\textbf{h}}_k^T} \right), {\text{Im}}\left( {{\textbf{h}}_k^T} \right), {\text{Re}}\left( {{\textbf{d}}_k^T} \right), {\text{Im}}\left( {{\textbf{d}}_k^T} \right)} \right]^T}, \nonumber
\end{align} 
where $\textbf{d}_\text{b}$ is the vectorized form of Bob's cascaded channel, with $\textbf{d}_\text{b} = \text{vec} ( \text{diag} \{ \textbf{f}^H_\text{b} \}  \textbf{G} )$. $\textbf{d}_{k}$ is analogous to $\textbf{d}_\text{b}$ in meaning. To effectively capture the collective influence of the STAR-IRS on the communication environment, the feature vector for the STAR-IRS node is formulated by averaging the direct and cascaded channels across all nodes, which is given by
\begin{align}
\bar{\textbf{x}} = \left[
\text{Re}(\bar{\textbf{h}}^T), \text{Im}(\bar{\textbf{h}}^T), \text{Re}(\bar{\textbf{d}}^T), \text{Im}(\bar{\textbf{d}}^T)\right]^T, \nonumber
\end{align} 
where $\bar{\textbf{h}}$ is the mean of $\textbf{h}_\text{b}$ and all $\textbf{h}_k$, and $\bar{\textbf{d}}$ is the mean of $\textbf{d}_\text{b}$ and all $\textbf{d}_k$. Thus, the node feature matrix is constructed as 
\begin{align}
\textbf{X}_{\text{input}} = 
\begin{bmatrix}
\text{Re}(\bar{\textbf{h}}^T) & \text{Im}(\bar{\textbf{h}}^T) & \text{Re}(\bar{\textbf{d}}^T) & \text{Im}(\bar{\textbf{d}}^T)\\
\text{Re}(\textbf{h}_\text{b}^T) & \text{Im}(\textbf{h}_\text{b}^T) & \text{Re}(\textbf{d}_\text{b}^T) & \text{Im}(\textbf{d}_\text{b}^T) \\
\text{Re}(\textbf{h}_1^T) & \text{Im}(\textbf{h}_1^T) & \text{Re}(\textbf{d}_1^T) & \text{Im}(\textbf{d}_1^T) \\
\text{Re}(\textbf{h}_2^T) & \text{Im}(\textbf{h}_2^T) & \text{Re}(\textbf{d}_2^T) & \text{Im}(\textbf{d}_2^T) \\
		\vdots & \vdots  & \vdots  & \vdots \\
\text{Re}(\textbf{h}_K^T) & \text{Im}(\textbf{h}_K^T) & \text{Re}(\textbf{d}_K^T) & \text{Im}(\textbf{d}_K^T)
\end{bmatrix}. \nonumber
\end{align}  
Since only Bob and the STAR-IRS are related to the generation of $\textbf{w}$, $\boldsymbol{\Omega}_\text{r}$, and $\tilde{\boldsymbol{\Omega}}_\text{t}$, the adjacency matrix is given by
\begin{align}
	\textbf{A} =
	\begin{bmatrix}
		0 & 1 & 0 &  0 & \cdots & 0 \\
		1 & 0 & 0 &  0 & \cdots & 0 \\
		1 & 1 & 0 &  0 & \cdots & 0 \\
		\vdots & \vdots  & \vdots  & \vdots & \ddots & \vdots \\
		1 & 1 & 0 &  0 & \cdots & 0 \\
        1 & 1 & 0 &  0 & \cdots & 0 \\
	\end{bmatrix}. \nonumber
\end{align} 
\subsection{Neural Network Architecture}
    \begin{figure}[!t]
		\centering
    \includegraphics[width=3.5in]{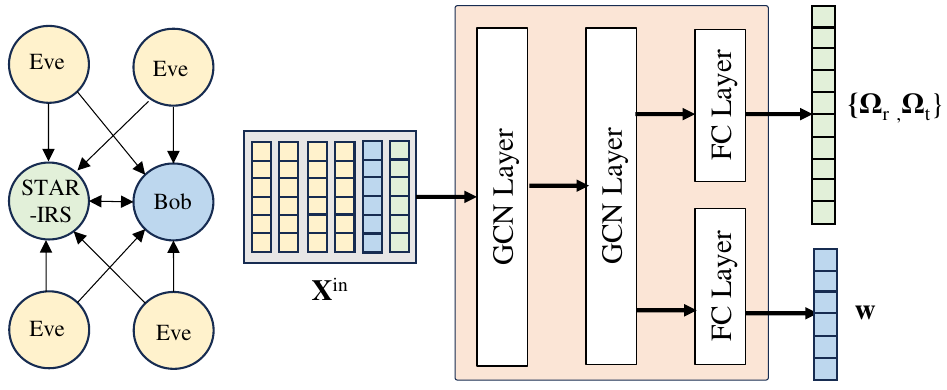}		\caption{Interrelationships between entities and neural network architecture.}
		\label{Fig2}
	\end{figure}
The proposed GNN architecture, illustrated in Fig. \ref{Fig2}, is carefully structured to effectively solve the formulated optimization problem (P1) by exploiting the underlying relationships within the indoor communication network. It comprises two graph convolutional network (GCN) layers, which capture complex interactions and dependencies among the STAR-IRS, Bob, and Eve nodes through iterative feature extraction and propagation across the graph structure. These GCN layers are followed by two independent fully connected (FC) layers, specifically designed to output the optimization variables crucial for enhancing communication security: the beamforming vector $\textbf{w}$, and the STAR-IRS reflection and transmission coefficient matrices $\{\boldsymbol{\Omega}_\text{r}, \tilde{\boldsymbol{\Omega}}_\text{t}\}$.

The first GCN layer is initialized with $2N+2NL$ input neurons, corresponding precisely to the dimensionality of the node features matrix. Subsequently, these features are propagated through each GCN layer by an aggregation-combination process. During the aggregation phase, each node collects and aggregates features from its neighboring nodes using a normalized adjacency matrix, mathematically represented as
\begin{equation}
	\textbf{H}^{(i+1)'} =  \tilde{\textbf{A}} \textbf{H}^{(i)} \textbf{F}^{(i)}, \nonumber
\end{equation}
where \( \textbf{H}^{(i)} \) denotes the node feature matrix at the $i$-th layer, and \( \textbf{F}^{(i)} \) denotes the learnable weight matrix of the same layer. The normalized adjacency matrix \( \tilde{\textbf{A}} \), which ensures efficient information flow by incorporating self-connections, is defined as
\begin{equation}
	\tilde{\textbf{A}} = \textbf{D}^{-\frac{1}{2}} (\textbf{A} + \textbf{I}) \textbf{D}^{-\frac{1}{2}},
\end{equation}
where \( \textbf{D} \) represents the degree matrix with each diagonal element $\textbf{D}_{ii}$ equals the sum of the adjacency matrix’s corresponding row elements $\textbf{A}_{ij}$.
Following the aggregation step, features undergo a linear transformation and then pass through a non-linear activation function, such as a rectified linear unit (ReLU). The nonlinear activation facilitates capturing complex relationships and interactions between different nodes. Mathematically, this combination operation is expressed as
\begin{equation}
	\textbf{H}^{(i+1)} = f(\textbf{H}^{(i+1)'}), \nonumber
\end{equation}
where $f(\cdot)$ denotes the activation function.
By sequentially stacking two GCN layers, the neural network effectively captures hierarchical interactions and dependencies within the graph structure. The extracted high-level features are then processed by two separate FC layers, specifically designed to enforce the constraints required by the optimization problem. The first FC layer outputs the STAR-IRS reflection and transmission coefficients  $\{\boldsymbol{\Omega}_\text{r}, \tilde{\boldsymbol{\Omega}}_\text{t}\}$, ensuring compliance with energy conservation and phase shift constraints. The second FC layer computes the beamforming vector $\textbf{w}$, maintaining adherence to the total transmit power constraint. This modular and hierarchical structure guarantees both effective learning and precise satisfaction of optimization constraints, ensuring secure and robust indoor communications.
\subsection{Output}
The whole neural network produces two tensors as its final output. To be specific, the first row of the output from the second GCN layer serves as the input to the initial FC layer to generate the reflection energy coefficient $\beta^\text{r}_l$, the cosine of the reflection phase $cos(\theta^\text{r}_l)$ and the cosine of the transmission phase $cos(\tilde{\theta}^\text{t}_l)$ for each components of the STAR-IRS. Note that the sigmoid activation function is applied before outputting $\beta^\text{r}_l$, $cos(\theta^\text{r}_l)$ and $cos(\tilde{\theta}^\text{t}_l)$. Mathematically, the output is given by 
\begin{align}
\textbf{v} =  [\beta^\text{r}_1, \beta^\text{r}_2, \cdots, \beta^\text{r}_L,  & cos(\theta^\text{r}_1), cos(\theta^\text{r}_2), \cdots, cos(\theta^\text{r}_L),  \nonumber\\
& cos(\tilde{\theta}^\text{t}_1), cos(\tilde{\theta}^\text{t}_2), \cdots, cos(\tilde{\theta}^\text{t}_L) ]. \nonumber
\end{align} 
The second row is fed into the second FC layer to produce the beamforming vector $\textbf{w}$ at Alice. \par
It is worth mentioning that the entire neural network architecture is readily transferable to the conventional STAR-IRS-aided secure communication scheme as given in Section V without any modifications. In this scheme, however, $\gamma_{k}$ needs to be modified to  
\begin{align}
\gamma_{k} =\frac{ | (\textbf{h}_k^H  + \textbf{f}_k^H \boldsymbol{\Omega}_\text{t}\textbf{G} ) \textbf{w}|^2 }{ \sigma_k^2 },~ k \in \mathcal{K}. \nonumber 
\end{align}
For the IRS-aided secure communication scheme as given in Section V, the GNN architecture requires some modification to the output of the first FC layer. To be specific, the output is given by 
\begin{align}
\textbf{v} =  [& cos(\theta^\text{r}_1), cos(\theta^\text{r}_2), \cdots, cos(\theta^\text{r}_L)]. \nonumber
\end{align} 
In this scheme, $\beta^\text{r}_l$ = 1 and $\beta^\text{t}_l$ = 0 are set.

\subsection{Training and Inference}
The GNN model undergoes unsupervised training, where its weights and biases are iteratively adjusted according to a loss function formulated as
\begin{align}
\mathcal{L} =  - \frac{\sum_{t=1}^{T}R_{s}^{(t)}}{T}. \nonumber
\end{align} 
$R_{s}^{(t)}$ denotes the secrecy rate of the indoor communication network for the $t$-th channel sample, with $T$ representing the total number of training samples. The loss function is minimized iteratively using stochastic gradient descent (SGD), gradually guiding the objective toward its optimum. Upon convergence, the GNN model effectively learns the complex interactions among Bob, Eve, and the STAR-IRS.\par
Although training the GNN model is time-consuming, it is performed offline and thus does not significantly affect system performance. By contrast, the complexity of online inference is critical for the model's practical deployment in communication systems. Algorithm 1 presents the procedure for implementing the GNN-based beamforming scheme. In the GNN, computational complexity is mainly dominated by the two GCN layers. For dense graphs, a single GCN layer has complexity $\mathcal{O}(C^2 + C F F')$, where \( C \) is the number of nodes, \( F \) is the number of input features per node, and \( F' \) is the number of output features per node. For sparse graphs, this can be more accurately expressed as $\mathcal{O}(M + C F F')$, with $M$ denoting the number of edges. Therefore, the online inference complexity of the GNN is compatible with the real-time requirements of communication systems.

\begin{algorithm}[t]
\caption{GNN-based beamforming scheme}
\begin{algorithmic}[1]
\State \textbf{Input}: Feature vectors $\textbf{x}_\text{b}$ for Bob and $\textbf{x}_k$ for all Eves, adjacency matrix \textbf{A}, parameters $(N,L,K)$ 
\State \textbf{Output}: Beamforming vector $\textbf{w}$, and STAR-IRS reflection and transmission coefficient matrices $\{\boldsymbol{\Omega}_\text{r}, \tilde{\boldsymbol{\Omega}}_\text{t}\}$.
\State $\textbf{x}_\textbf{w} \gets \text{concat} (\textbf{x}_\text{b}; \textbf{x}_k)$ 
\State $\textbf{X}_{\text{input}} \gets \text{concat} (\text{mean}(\textbf{x}_\textbf{w}); \textbf{x}_\textbf{w})$
\State $\textbf{X} \gets \text{ReLU}(\text{GCN1}(\textbf{X}_{\text{input}}, \textbf{A}))$
\State $\textbf{X} \gets \text{ReLU}(\text{GCN2}(\textbf{X}, \textbf{A}))$
\State $\textbf{v}_\text{tmp} \gets \sigma(\text{FC}_\textbf{v}(\textbf{X}[0, :]))$
\State $\boldsymbol{\beta}_\text{r} \gets \textbf{v}_\text{tmp}[0:L]$
\State $\boldsymbol{\theta}_\text{r} \gets \textbf{v}_\text{tmp}[L:2L] + j\sqrt{1-\textbf{v}_\text{tmp}[L:2L]^2}$
\State $\boldsymbol{\Omega}_\text{r} \gets \text{diag} \{ \sqrt{\boldsymbol{\beta}_\text{r}} \boldsymbol{\theta}_\text{r} \}$
\State $\boldsymbol{\beta}_\text{t}  \gets 1-\boldsymbol{\beta}_\text{r}$
\State $\boldsymbol{\theta}_\text{t} \gets \textbf{v}_\text{tmp}[2L:3L] + j\sqrt{1-\textbf{v}_\text{tmp}[2L:3L]^2}$
\State $\tilde{\boldsymbol{\Omega}}_\text{t} \gets \text{diag} \{\sqrt{\boldsymbol{\beta}_\text{t}} \boldsymbol{\theta}_\text{t} \}$
\State $\textbf{w} \gets \sqrt{P} \, \text{LayerNorm} \left( \textbf{X}[1, :] \right)$
\end{algorithmic}
\end{algorithm}
While the GNN model's training is time-consuming, it occurs offline and thus has little effect on performance. In contrast, the complexity of the online inference process is critical in determining whether the model can be applied to the communication system. In the GNN model, the computational complexity is primarily determined by the two GCN layers. The complexity of a single GCN layer is given by $\mathcal{O}(C^2 + C F F')$ for dense graphs, where \( C \) is the number of nodes, \( F \) is the number of input features per node, \( F' \) is the number of output features per node. For sparse graphs, the complexity can be more accurately represented as $ \mathcal{O}(M + C F F')$, where $M$ denotes the number of edges. Hence, the complexity of the online inference process in the GNN model can meet the real-time demands of communication systems. 
\section{Extension to Imperfect Eavesdropping CSI}\label{sec:extension}
This section extends the study to scenarios with imperfect eavesdropping CSI. Specifically, the direct channel vector $\textbf{h}_{k}$ from Alice to the $k$-th Eve and the corresponding cascaded channel $\textbf{d}_{k} = \text{diag}\{ \textbf{f}^H_k \} \textbf{G}$ can be expressed as
\begin{align}
\textbf{h}_{k} &= \bar{\textbf{h}}_{k}+\hat{\textbf{h}}_{k}, \nonumber 
\end{align} 
and
\begin{align}
\textbf{d}_{k} &= \bar{\textbf{d}}_{k} + \hat{\textbf{d}}_{k}, \nonumber
\end{align}
where $\bar{\textbf{h}}_{k}$ and $\bar{\textbf{d}}_{k}$ denote the estimated CSI of the direct channel and the cascaded channel, respectively, while $\hat{\textbf{h}}_{k}$ and $\hat{\textbf{d}}_{k}$ represent the corresponding channel estimation errors. Assuming that the channel estimation errors follow a Gaussian distribution, they can be mathematically expressed as
\begin{align}
\hat{\textbf{h}}_{k} \sim \mathcal{CN}\left(\textbf{0}, \sigma_{h}^{2}\textbf{I}\right), \nonumber
\end{align} 
and
\begin{align}
\hat{\textbf{d}}_{k} \sim \mathcal{CN}\left(\textbf{0}, \sigma_{d}^{2}\textbf{I}\right), \nonumber
\end{align}
where $\sigma_{h}^{2}$ and $\sigma_{d}^{2}$ denote the variances of the channel estimation errors for the direct and cascaded channels, respectively.
Based on this, the signal received at the $k$-th Eve can be rewritten as
\begin{align}
y_k = (\bar{\textbf{h}}_{k}+\hat{\textbf{h}}_{k}) \textbf{w} s + \tilde{\boldsymbol{\omega}}_\text{t}^H (\bar{\textbf{d}}_{k}+ (\hat{\textbf{d}}_{k})^H ) \textbf{w} z + n_k. \nonumber
\end{align}
Despite the imperfect CSI, the maximum SINR that Eve can achieve is still determined by \eqref{eq2}. This is because, although Bob cannot perfectly obtain Eve's SINR, Eve herself has access to it. Therefore, to ensure secure communication, it is necessary to consider the worst eavesdropping scenario. However, the instantaneous secrecy rate should be substituted by its expected value, which is given by
\begin{equation} 
\bar{R}_{s} = \mathbb{E}_{\hat{\textbf{h}}_{k}, \hat{\textbf{d}}_{k}} \left\{ \left[ \log_2\left ( 1 + \gamma_\text{b} \right) - \max_{\forall k \in \mathcal{K}} ~\log_2\left ( 1 + \gamma_k \right) \right]^+ \right\}. \nonumber
\end{equation}
In the presence of channel uncertainty, the objective is reformulated as maximizing the expected secrecy rate, leading to the following optimization problem:
\begin{equation}\label{P2}\nonumber
	\begin{aligned}
		(\text{P2})\quad \mathop{\max}_{\textbf{w}, \boldsymbol{\Omega}_\text{r},\tilde{\boldsymbol{\Omega}}_\text{t}} \quad & \bar{R}_{s} \\
		s.t. \quad 
		&\text{C1} - \text{C6}, \\
	\end{aligned}
\end{equation}
When CSI is imperfect, the GNN-based beamforming method can continue to function by incorporating Bob’s full CSI along with the estimated CSI of Eve as the input feature matrix, while redesigning the loss function to optimize the average secrecy rate.
\section{FPGA-Based Acceleration}
To satisfy the strict latency requirements of real-time beamforming in the considered STAR-IRS assisted secure indoor communication system, we develop a dedicated FPGA-based accelerator for the inference of the trained GNN model. Unlike GPU or CPU implementations, which are often constrained by sequential execution and memory bandwidth, the proposed FPGA design leverages fine-grained parallelism and pipelined computation to accelerate processing. The hardware microarchitecture consists of three primary processing modules: GCN convolution layers, FC layers, and normalization units.
\subsection{GCN Layer Implementation}
Each GCN convolution layer is implemented as a pipelined matrix-vector multiplication engine, allowing simultaneous computation of multiple output nodes. Nonlinear activation functions, specifically ReLU for hidden layers, are realized using hardware-efficient piecewise linear approximations or lookup tables. Complex-valued operations required for STAR-IRS phase-amplitude modulation are handled via separate real and imaginary channels, enabling parallel processing of multiple elements. The convolution layers are executed in a streaming dataflow fashion to fully exploit pipeline parallelism.

\subsection{FC Layer and Normalization}
The FC layers for predicting the transmitter’s beamforming vector as well as the STAR-IRS’s transmission and reflection matrices are implemented as pipelined matrix-vector multiplication units. Post-processing operations, such as normalization and amplitude-phase decomposition, are incorporated into the pipeline to ensure that the resulting vectors satisfy the physical constraints of the STAR-IRS. Batch processing is supported by interleaving multiple input feature vectors across pipeline stages. Critical inner loops in matrix multiplications and vector normalizations are unrolled to reduce initiation interval (II) to one cycle, maximizing throughput.

\subsection{Memory and Dataflow Optimization}
Memory access patterns are carefully optimized to reduce latency and off-chip memory traffic. Weights and biases are stored in on-chip BRAMs, while input and output feature streams are aligned to support burst transfers. Streaming feature vectors through the pipeline minimizes intermediate memory writes. Fixed-point arithmetic with carefully chosen word lengths strikes a balance between precision and resource utilization, ensuring that the predicted beamforming vectors and matrices maintain high fidelity for secure and efficient communication.
\subsection{Performance and Scalability}
The proposed FPGA-based design provides low-latency and energy-efficient computation for GNN inference in wireless beamforming. The architecture is scalable with respect to the number of antennas and STAR-IRS elements. HLS-based design methodology allows rapid prototyping and optimization for different hardware targets. By combining parallelism, pipelining, and memory optimization, the accelerator achieves significant speedup compared to CPU or GPU implementations, making it suitable for real-time deployment the considered STAR-IRS assisted secure indoor communication systems.
\section{Simulation and Experimental Results}
\begin{figure}[!t]
    \centering
\includegraphics[width=3.0in]{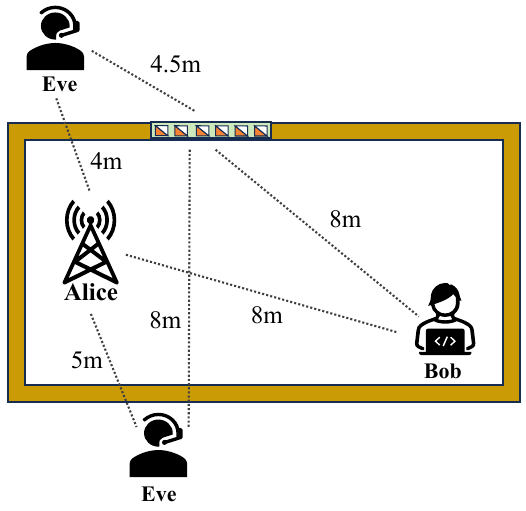}
    \caption{A random observation of the spatial positions of Alice, Bob, and Eves.}
    \label{Fig3}
\end{figure}
\begin{table}[t]
\scriptsize
\begin{center}
\caption{Simulation Parameters.}\label{T1}~~\\
\begin{tabular}{c|c|c} 
\hline
\textbf{Notation}        &  \textbf{Description}              & \textbf{Value}                   \\ \hline\hline
$d_\text{ab}$            &  Distances from Alice to Bob       & $8$ m                            \\ \hline
$d_\text{sb}$            &  Distances from STAR-IRS to Bob    & $8$ m                            \\ \hline
$d_\text{ae}$            &  Distances from Alice to Eve       & $[4~{\rm m}, 8~{\rm m}]$         \\ \hline
$d_\text{se}$            &  Distances from STAR-IRS to Eve    & $[4~{\rm m}, 8~{\rm m}]$         \\ \hline
$N$                      &  Number of antennas at Alice       & $8$                              \\ \hline
$P$                      &  Alice’s transmit power            & $18~{\rm dBm}$                   \\ \hline
$L$                      &  Number of elements at STAR-IRS    & $80$                             \\ \hline
$K$                      &  Number of Eves                    & $2$                              \\ \hline
$\kappa$                 &  Rician factor                     & 0.3                              \\ \hline
$d_0$                    &  Reference distance                & $1~{\rm m}$                 \\ \hline
$\sigma^2$               &  Noise variance at Bob and Eves   & $-90~{\rm dBm}$             \\ \hline
\end{tabular}
\end{center}
\end{table}
This section evaluates the improvements in SRM achieved by the proposed STAR-IRS-enabled secure communication strategy and the adopted GNN-based optimization scheme for indoor environments. In addition, the computational performance of the corresponding FPGA-based GNN accelerator is assessed.
\subsection{Simulation Setup}
In the simulations, the Alice-Bob, Alice-Eve, STAR-IRS-Bob, STAR-IRS-Eve and Alice-STAR-IRS channels are modeled as Rician fading with a Rician factor of $\kappa$ = 0.3. The channel path loss is given by $PL = PL_0 - 25 \log_{10} \left( \frac{d}{d_0} \right)$ \text{dB}, where $d$ is the transmission distance and $d_0$ = 1 m denotes the reference distance. The distances from Alice to Bob, from the STAR-IRS to Bob, and from Alice to the STAR-IRS are all set to 8 m. Fig. \ref{Fig3} shows a random observation of the spatial positions of Alice, Bob, and Eves. In contrast, the distances from Alice to Eve and from the STAR-IRS to Eve are randomly generated within the range of 4 m to 8 m, reflecting the realistic variability in the spatial locations of potential eavesdroppers in the considered environment and capturing diverse propagation conditions that may affect the secrecy performance of the communication system. Some other parameters are set as follows. Alice has \( N = 8 \) antennas, there are \( K = 2 \) Eves, the STAR-IRS consists of \( L = 80 \) elements, Alice's transmit power is \( P = 18 \) dBm, and the noise variance at Bob and Eves is $-90~{\rm dBm}$. These parameters can vary when used as the horizontal axis. The simulation parameters for the considered STAR-IRS assisted secure indoor communication system, are summarized in Table \ref{T1}. All the simulated schemes are detailed below. 
\begin{itemize}
    \item \texttt{AN-GNN}: This legend denotes the proposed STAR-IRS-enabled secure strategy, where the transmitted signal at the STAR-IRS undergoes random phase change at the symbol level, transforming into AN during transmission. Moreover, the GNN-based scheme is applied to settle the optimization problem (P1).
    
    \item \texttt{CONV-GNN}: This legend denotes the conventional STAR-IRS-enabled secure strategy. Unlike \texttt{AN-GNN}, the transmitted signal is used directly to suppress Eve's signal reception without random phase modulation.
    
    \item \texttt{IRS-GNN}: This legend denotes the reflection-only secure strategy, where the STAR-IRS operates solely on the reflection function.

    \item \texttt{AN-MRT}: This legend denotes the maximum ratio transmission (MRT) scheme \cite{MRT} instead of the GNN-based one. The STAR-IRS's reflection and transmission coefficients are configured randomly, with uniform distribution of reflected and transmitted energy.

    \item \texttt{AN-ZF}: This legend denotes the zero forcing (ZF) scheme \cite{ZF} instead of the GNN-based scheme. The STAR-IRS's reflection and transmission coefficients are configured randomly, with even distribution of reflected and transmitted energy.
    
    \item \texttt{AN-MMSE}: This legend represents the minimum mean square error (MMSE) scheme \cite{MMSE} in place of the GNN-based one. The STAR-IRS's reflection and transmission coefficients are configured randomly, with even distribution of reflected and transmitted energy.
\end{itemize}
\subsection{Simulation Results}
\begin{figure}[!t]
    \centering
\includegraphics[width=3.7in]{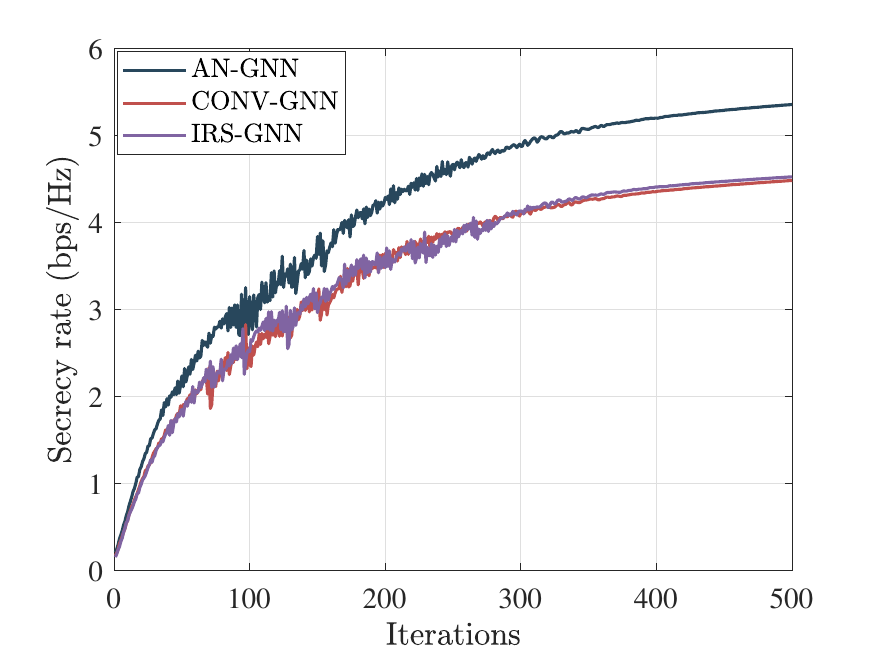}
    \caption{Convergence behaviors of the GNN-based scheme for \texttt{AN-GNN}, \texttt{CONV-GNN}, and \texttt{IRS-GNN}.}
    \label{Fig4}
\end{figure}
Fig. \ref{Fig4} illustrates the convergence trajectory of the GNN-based scheme for \texttt{AN-GNN}, \texttt{CONV-GNN}, and \texttt{IRS-GNN}. It can be observed that all three schemes converge rapidly during training, reaching a relatively stable secrecy rate within approximately 300 iterations. The rapid convergence indicates that the GNN effectively learns the complex interactions among Bob, Eves, and the STAR-IRS, efficiently capturing the mapping from channel states to optimal beamforming and STAR-IRS coefficients. After around 300 iterations, the secrecy rate continues to improve slightly but at a much slower pace, suggesting that the model has already captured the dominant patterns in the data. The marginal improvements beyond this point reflect fine-tuning of the network weights, which further refines the optimization of beamforming and STAR-IRS coefficients, but with diminishing returns. Notably, \texttt{AN-GNN} consistently achieves a higher secrecy rate compared to \texttt{CONV-GNN} and \texttt{IRS-GNN} throughout the training process, highlighting the effectiveness of incorporating symbol-level random phase modulation together with the GNN-based joint optimization.\par
\begin{figure}[!t]
    \centering
\includegraphics[width=3.7in]{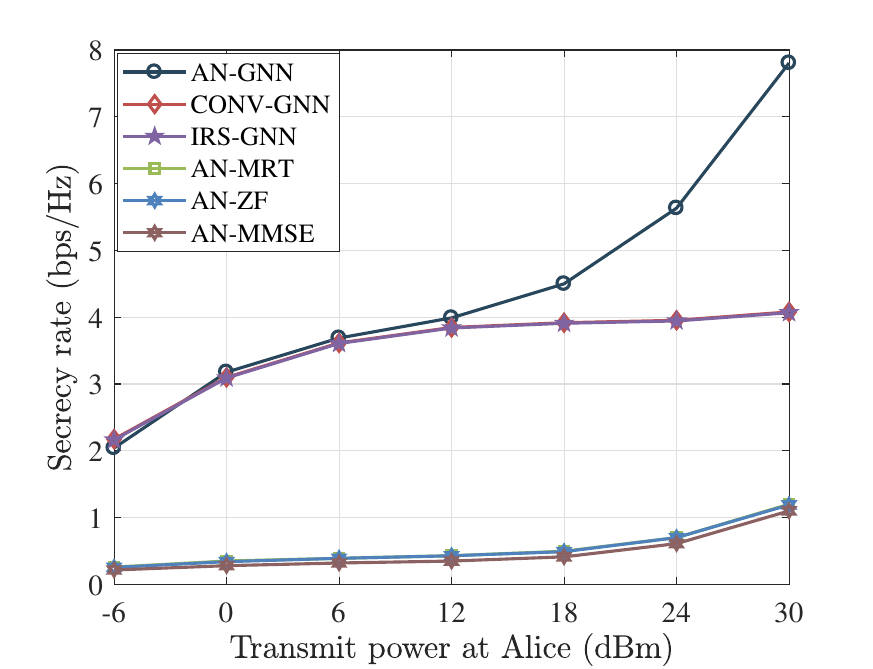}
    \caption{The secrecy rate \textit{vs.} the transmit power at Alice.}
    \label{Fig5}
\end{figure}
Fig. \ref{Fig5} compares the secrecy rate achieved by different approaches as a function of the transmit power at Alice. We observe that higher transmit power generally enhances secrecy performance. When the transmit power at Alice is not large enough,
the performance difference between \texttt{AN-GNN}, \texttt{CONV-GNN} and \texttt{IRS-GNN} is very small. Conversely,
\texttt{AN-GNN} significantly outperforms both \texttt{CONV-GNN} and \texttt{IRS-GNN} at the high transmit power. The reason is that
the impact of noise on SINR decreases with the transmit power increasing and AN is generated in the scheme of \texttt{AN-GNN}.
This result confirms the advantage of modulating the transmitted signal with random symbol-level phase shifts to degrade the signal quality at Eves. On the other hand, the GNN-based scheme demonstrates superior performance compared to \texttt{MRT}, \texttt{ZF}, and \texttt{MMSE}, illustrating the effectiveness of the GNN approach. \par
\begin{figure}[!t]
	\centering
	\includegraphics[width=3.7in]{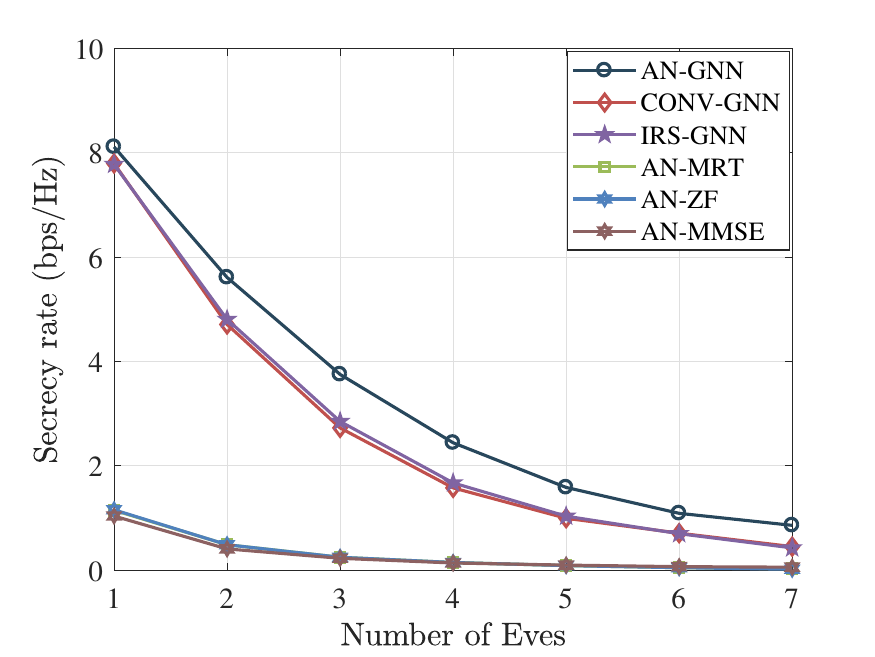}
	\caption{The secrecy rate \textit{vs.} the number of Eves.}
	\label{Fig6}
\end{figure}
Fig. \ref{Fig6} illustrates how the secrecy rate varies as the number of Eves increases. The analysis demonstrates that a larger population of Eves negatively impacts the achievable secrecy rate. This trend is intuitive, as a greater number of Eves result in increased spatial degrees of freedom, thereby raising the maximum eavesdropping rate. Among the schemes, \texttt{AN-GNN} consistently achieves the highest secrecy rate across all Eve counts, followed closely by \texttt{IRS-GNN}, while \texttt{CONV-GNN} exhibits the most pronounced decline. The robustness of \texttt{AN-GNN} in multi-Eve scenarios can be attributed to its artificial noise strategy, which effectively disrupts the eavesdroppers’ ability to decode the signal, even as their numbers grow. Conversely, the steeper drop in \texttt{CONV-GNN}’s performance suggests that its conventional convolutional techniques may struggle to adapt to the increased complexity introduced by multiple Eves, emphasizing the value of adaptive noise modulation for maintaining secrecy.\par
\begin{figure}[!t]
    \centering
\includegraphics[width=3.7in]{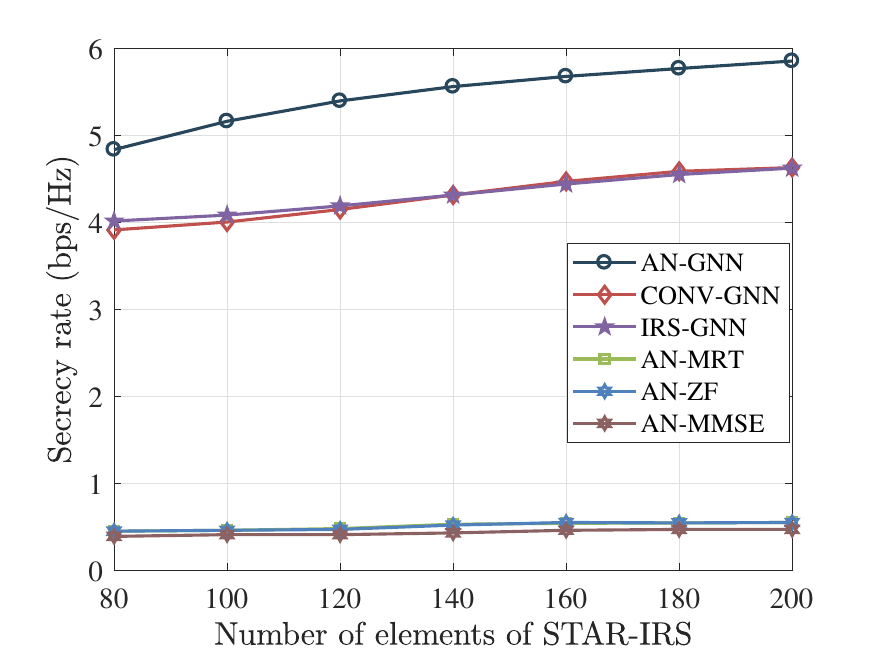}
    \caption{The secrecy rate \textit{vs.} the number of elements of the STAR-IRS.}
    \label{Fig7}
\end{figure}
Fig. \ref{Fig7} illustrates how the secrecy rate varies with the number of elements of the STAR-IRS. It is observed that the secrecy rate increases as the number of STAR-IRS elements grows across all schemes. This enhancement is due to the increased reflected or transmitted power and the greater spatial degrees of freedom provided by additional elements, which improve beamforming and interference suppression capabilities. Among the schemes, \texttt{AN-GNN} consistently achieves the highest secrecy rate, displaying a near-linear increase, which indicates its effective utilization of the additional elements to enhance the signal for the Bob while simultaneously degrading it for Eves. The superior performance of \texttt{AN-GNN} suggests that its integration of AN synergizes with the spatial advantages of a larger STAR-IRS, optimizing the balance between signal enhancement and interference generation.\par
\begin{figure}
\centering
\includegraphics[width= 3.7in]{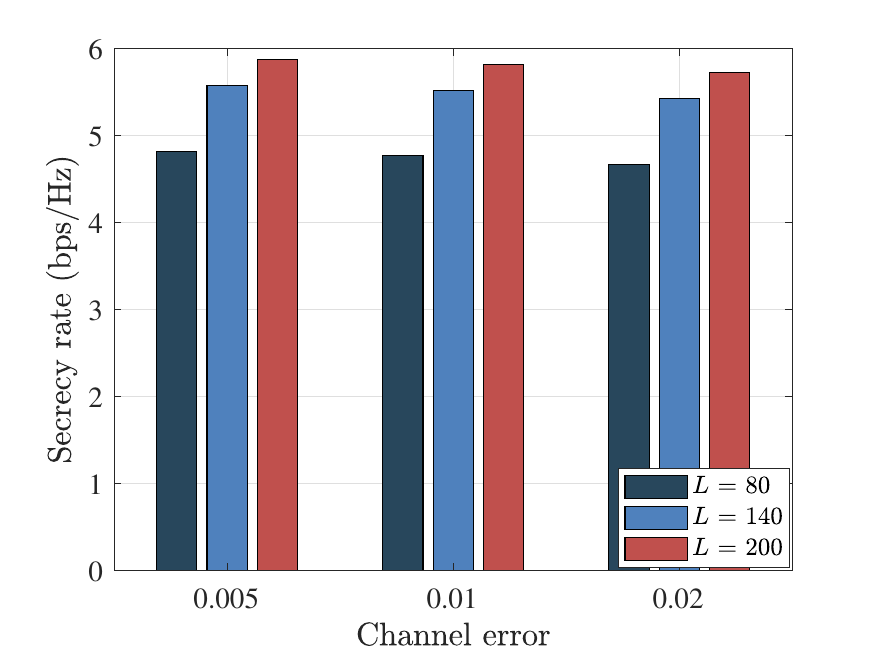}
\caption{The influence of imperfect eavesdropping CSI on the achievable communication performance of the GNN-based optimization scheme, assuming Gaussian-distributed channel errors.}
\label{Fig8}
\end{figure}

Fig. \ref{Fig8} depicts the influence of imperfect eavesdropping CSI on the achievable secrecy performance of the GNN-based optimization scheme, assuming Gaussian-distributed channel estimation errors. The $x$-axis represents the mean square error of the estimated channels. As the channel estimation error increases, a slight degradation in the achievable secrecy rate is observed. Nevertheless, the proposed GNN-based method demonstrates strong robustness to channel uncertainty, maintaining a high secrecy rate even under relatively large estimation errors. This result indicates that the GNN can effectively learn to optimize the beamforming vector and STAR-IRS coefficients using only the estimated CSI, mitigating the negative impact of imperfect knowledge of the eavesdroppers’ channels.
\subsection{FPGA-based GNN Accelerator}
\begin{figure}[!t]
    \centering
\includegraphics[width=3.7in]{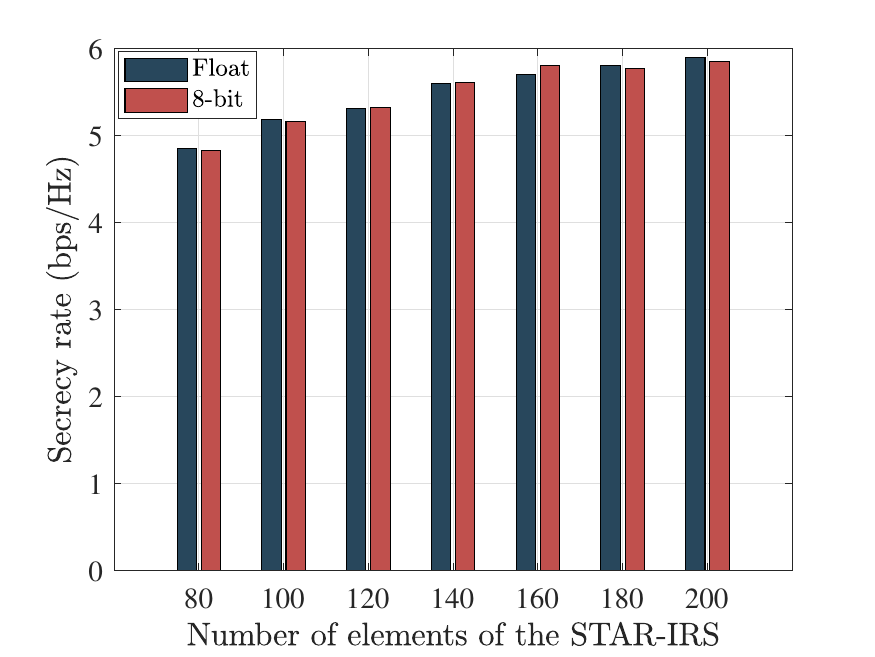}
    \caption{Comparison of communication performance achieved by GNN-based optimization before and after quantization.}
    \label{Fig9}
\end{figure}

This section presents an experimental analysis of the FPGA-based GNN accelerator, where the experiment is conducted on a Xilinx Zynq-7000 XC7Z100-2FFG900 SoC FPGA. The accelerator implements fixed-point arithmetic on FPGA to minimize hardware costs and energy usage. While this approach significantly improves resource efficiency, it comes with a marginal compromise in neural network inference precision, as demonstrated in Fig. \ref{Fig9}. Regardless of the number of STAR-IRS elements, the quantized inference results remain highly consistent with those obtained using floating-point representation. The measured latency is 8.992 ms with a 10 ns clock period. These results demonstrate that the proposed GNN-based optimization framework satisfies the latency requirements for most applications in the STAR-IRS aided indoor wireless communication system.
\section{Conclusions}
This paper proposed using random symbol-level phase shifts at the STAR-IRS to transform the transmitted signal into AN, enhancing the secrecy performance of indoor communications. Additionally, a GNN-based scheme was developed to jointly optimize the beamforming at Alice and the reflection coefficient and transmission coefficient matrices at the STAR-IRS. Simulation results demonstrated that the proposed STAR-IRS-based secure communication strategy provides superior secrecy performance compared to both the conventional STAR-IRS and reflection-only schemes. Moreover, the GNN-based scheme outperforms traditional approaches such as MRT, ZF, and MMSE. Experimental results validate that the FPGA-based GNN accelerator achieves low-latency inference.
	
	\ifCLASSOPTIONcaptionsoff
	\newpage
	\fi
	%

	%
	%
	%

	
	

\end{document}